\documentclass[referee]{raa}
\usepackage{graphicx,times}
\usepackage{natbib}
\usepackage{amssymb,amsmath}
\bibpunct{(}{)}{;}{a}{}{,}
\usepackage{float}

\usepackage[pagebackref=true]{hyperref}
\hypersetup{colorlinks = true, linkcolor = green, anchorcolor = red, citecolor = blue, filecolor = red, pagecolor = red, urlcolor = red}

\begin{document}

   \title{The decay process of an ${\rm \alpha}$-configuration sunspot}

 \volnopage{ {\bf 20XX} Vol.\ {\bf X} No. {\bf XX}, 000--000}
   \setcounter{page}{1}

   \author{Yang Peng\inst{1,2}, Zhi-Ke Xue\inst{1,3}, Xiao-Li Yan\inst{1,3}, Aimee A. Norton\inst{4}, Zhong-Quan Qu\inst{1}, Jin-Cheng Wang\inst{1,3}, Zhe Xu\inst{1,3}, Li-Heng Yang\inst{1,3}, Qiao-Ling Li\inst{6,7}, Li-Ping Yang\inst{1,2}, Xia Sun\inst{2,5}}

\institute{ Yunnan Observatories, Chinese Academy of Sciences, Kunming 650216, China; {\it pengyang@ynao.ac.cn\\}
\and
University of Chinese Academy of Sciences, Beijing 100049, China\\
\and
Yunnan Key Laboratory of Solar Physics and Space Science, Kunming 650216, China\\
\and
HEPL, Stanford University, Stanford, CA 94305, USA\\
\and
Yunnan normal University, Kunming 650500, China\\
\and
Department of Physics, Yunnan University, Kunming 650091, China
\and
Department of Astronomy, Key Laboratory of Astroparticle Physics of Yunnan Province, Yunnan University, Kunming 650091, China\\
\vs \no
   {\small Received 20XX Month Day; accepted 20XX Month Day}
}

\abstract{
	The decay of sunspot plays a key role in magnetic flux transportation in solar active regions (ARs). To better understand the physical mechanism of the entire decay process of a sunspot, an $\alpha$-configuration sunspot in AR NOAA 12411 was studied. Based on the continuum intensity images and vector magnetic field data with stray light correction from SDO/HMI, the area, vector magnetic field and magnetic flux in the umbra and penumbra are calculated with time, respectively. Our main results are as follows: (1) The decay curves of the sunspot area in its umbra, penumbra, and whole sunspot take appearance of Gaussian profiles. The area decay rates of the umbra, penumbra and whole sunspot are -1.56 $\rm MSH\ day^{-1}$, -12.61 $\rm MSH\ day^{-1}$ and -14.04 $\rm MSH\ day^{-1}$, respectively; (2) With the decay of the sunspot, the total magnetic field strength and the vertical component of the penumbra increase, and the magnetic field of the penumbra becomes more vertical. Meanwhile, the total magnetic field strength and vertical magnetic field strength for the umbra decrease, and the inclination angle changes slightly with an average value of about 20 degrees; (3) The magnetic flux decay curves of the sunspot in its umbra, penumbra, and whole sunspot exhibit quadratic patterns, their magnetic flux decay rates of the umbra, penumbra and whole sunspot are $-9.84\times10^{19}$ $\rm Mx\ day^{-1}$, $-1.59\times10^{20}$ $\rm Mx\ day^{-1}$ and $-2.60\times10^{20}$ $\rm Mx\ day^{-1}$, respectively. The observation suggests that the penumbra may be transformed into the umbra, resulting in the increase of the average vertical magnetic field strength and the reduction of the inclination angle in the penumbra during the decay of the sunspot.
}

   \authorrunning{Y. Peng et al. }            
   \titlerunning{The evolution of a decaying sunspot}  
   \maketitle
%
\section{Introduction}           
\label{sect:intro}
The decay of sunspot plays an important role in the magnetic field evolution of the active region \citep[AR,][]{Wallenhorst+Howard+1982, Wallenhorst+Topka+1982} and the magnetic flux transportation on the solar surface \citep{DeVore+etal+1984, Wang+etal+2002}. It also affects the solar radiation \citep{Willson+etal+1981, Lean+2013}. Thus, it is a hot topic of solar physics. Many early studys mainly studied the sunspot area decay law based on sunspot catalogs (e.g. Greenwich Photoheliographic Results and Mount Wilson sunspot data), and rarely included the decay law of the sunspot magnetic field \citep{Howard+1992, Lustig+Wohl+1995, Javaraiah+2011, Javaraiah+2012}. With the development of the observation equipment and spectral inversion technology, the magnetic field on the Sun's surface can be measured more precisely. Therefore, the decay laws of sunspots have been widely studied \citep{Norton+etal+2017, Benko+etal+2018, Verma+etal+2018, Li+etal+2021}. However, the evolution of sunspot from its peak period to disappearance is rarely reported, and the sunspot decay mechanism has not been well understood.

The area decay is an important aspect for a decaying sunspot. The decay law of the sunspot area is generally considered to be linear \citep{Chapman+etal+2003, Li+etal+2021}, while several observations indicate that it is quadratic \citep{Petrovay+etal+1997, Litvinenko+Wheatland+2015}. \cite{Li+etal+2021} studied the decay of eight $\alpha$-configuration sunspots, and discovered that the area decay of $\alpha$-configuration sunspots could be approximately linear, and the decay rate was not a constant. \cite{Benko+etal+2021} obtained a linear decay law for the area of umbra and a quadratic decay law for the penumbra respectively by analyzing the evolution of a decaying sunspot. \cite{Solanki+2003} proposed that over 90\% of decaying sunspots showed a linear decrease in area. A parabolic decay law was discovered by \cite{Litvinenko+Wheatland+2015}, and was supported by the research of \cite{Petrovay+Moreno-Insertis+1997}. Many studies show that different decay laws may correspond to different decay mechanisms \citep{Meyer+etal+1974, Krause+Ruediger+1975, Petrovay+Moreno-Insertis+1997, Litvinenko+Wheatland+2015, Xue+etal+2021}. The Umbra-Penumbra area ratio (U/P) of sunspot is also an important parameter during the decay of sunspot, which is tightly correlated with the magnetic field \citep{Jin+etal+2006}. \cite{Hoyt+Schatten+1998} suggested that the higher U/P value means the higher convective velocity and faster sunspot decays, since the convective velocity is proportional to the sunspot decay. \cite{Chapman+etal+2003} also discovered that the total area of sunspots and the U/P value are strongly correlated to the sunspot decay rate. \cite{MartinezPillet+etal+1993} concluded that the U/P value is nearly a constant.

The magnetic field is an important parameter for the sunspot decay. Many researches have shown that the magnetic field of the penumbra becomes more vertical at the beginning of the sunspot decay \citep{Watanabe+etal+2014, Bellot+etal+2008}. By studying the rapid decay of a penumbra after solar flares, \cite{Wang+etal+2004} discovered that the magnetic field in the sunspot inclined to be more vertical, and a part of penumbral magnetic field tend to convert into the umbral magnetic field. \cite{Verma+etal+2018} studied the decay process of AR NOAA 12597 and found that, a dark region filled with umbra dots was observed in the penumbral filaments. The velocity and magnetic properties of this dark region were similar to those of the umbra, and the magnetic field of the penumbra became more vertical. \cite{Jur+2011} selected the inner penumbra boundaries of 10 sunspots, and proved that the vertical magnetic field of the inner penumbra boundary was a constant. And then, \cite{Jur+etal+2018} reinforced this conclusion, and found that this constant value was almost 1867 G by analyzing 88 scans of 79 active regions observed with spectropolarimeter on board the Hinode satellite. \cite{Schmassmann+etal+2018} also supported the conclusion of \cite{Jur+2011} by analyzing vector magnetic field data for Active Region (AR) NOAA 11591 from Helioseismic and Magnetic Imager \citep[HMI,][]{Schou+etal+2012} on board Solar Dynamics Observatory \citep[SDO,][]{Scherrer+etal+2012}, but the constant was considered to be 1693 G. Whereas, \cite{Benko+etal+2018} discovered that the vertical magnetic field of the inner penumbra boundary of a decaying sunspot tended to decrease during the first 4 days and increases again during the last stages of the decay.

It is a common knowledge that the change of the magnetic flux acts a significant role in the decay process of sunspots. In general, the magnetic flux of decaying sunspots decreases linearly \citep{Verma+etal+2012, Rempel+2015}. \cite{Li+etal+2021} suggested that the magnetic flux of $\alpha$-configuration sunspots decrease linearly based on vector magnetic field data observed by SDO/HMI. And their decay rates are between $\rm -1.4\times 10^{20}$ $\rm Mx\ day^{-1}$ and $\rm -4.9\times 10^{20}$ $\rm Mx\ day^{-1}$. \cite{Sheeley+etal+2017} studied the magnetic flux decay of 36 sunspots, and found that the magnetic flux of some sunspots decreases linearly. Their decay rates are $\rm 2\sim 4\times 10^{20}$ $\rm Mx\ day^{-1}$. Some studies have shown that the emergence of the magnetic flux around sunspots has an important influence on the formation \citep{Schlichenmaier+etal+2010, Rezaei+etal+2012} and decay \citep{Verma+etal+2018} of the penumbra. Additionally, the moving magnetic features (MMFs) are usually observed in the vicinity of decaying sunspots, and it transported magnetic flux to surrounding network during the sunspot decay \citep{Verma+etal+2012, Deng+etal+2007}.

To better understand the whole decay mechanism of sunspots, the evolution of an $\alpha$-configuration sunspot is studied in this paper. Observations and Methods are described in section \ref{sect:obs}. In section \ref{sect:result}, the results of the sunspot decay are analyzed. At the end, conclusions and discussion are shown in section \ref{sect:conclusion}.
 
\section{Observations and Methods}
\label{sect:obs}
 The basic information on the decay process of an $\alpha$-type sunspot is listed in the Table \ref{sunspot information}, which includes the AR number of the sunspot, beginning and ending time of the chosen data, and the position of the sunspot, respectively.
\begin{table}[ht]
\centering
\caption{Sunspot information}
	\begin{tabular}{ccccc}
		\hline		
		AR number & Start date & Start position & End date & End position \\ \hline
		NOAA 12411     &20150907 00:00UT &    N$13^{\circ}$E$39^{\circ}$      &20150912 06:00UT&    N$15^{\circ}$W$29^{\circ}$          \\ \hline
	\end{tabular}

\label{sunspot information}
\end{table}

The analysis is based on the data observed with SDO/HMI. The hmi.sharp\_cea\_720s data is a series of Space-weather HMI Active Region Patches (SHARP) data taken by SDO/HMI. Its temporal cadence and pixel scale are 720s and $0.5^{"}$, respectively \citep{Bobra+etal+2014, Hoeksema+etal+2014}. The hmi.sharp\_cea\_720s\_dconS data is based on the hmi.sharp\_cea\_720s data with stray light correction\footnote{http://jsoc.stanford.edu/doc/data/hmi/PSF\_corrected.html}. The temporal cadence and pixel scale of the hmi.sharp\_cea\_720s\_dconS data are the same as that of the hmi.sharp\_cea\_720s data. The continuum intensity images and vector magnetic field data from the hmi.sharp\_cea\_720s\_dconS data are selected to study the decay process of the sunspot. The temporal cadence of the selected data is set to be 1 hour. The vector magnetic field data are obtained by inversion from six sample points on the Fe I 6173.3 Å spectral line. The vector magnetic field data consists of three components, namely, $B_{R}$, $B_{P}$, $B_{T}$. The $B_{P}$, $B_{T}$, $B_{R}$ represent $\phi$ (westward) component of the Cylindrical Equal-Area (CEA) projection vector magnetic field in the direction of solar rotation, $\theta $ (southward) component of the CEA vector magnetic field, radial (out of photosphere) component of the CEA vector magnetic field, respectively \citep{Hoeksema+etal+2014}.

The umbra and penumbra of the sunspot on the photosphere are determined through the following process: First, an average continuum  intensity of the solar quiet region around the sunspot is calculated as $I_{0}$. Next, the intensities of continuum images are normalized by $I_{0}$. The umbra and penumbra are determined according to following two criteria respectively: $I_{umbra}\leqslant 0.55I_{0}$ and $0.55I_{0}<I_{penumbra}\leqslant 0.94I_{0}$, where $I_{umbra}$ and $I_{penumbra}$ are continuum intensity values of the umbra and penumbra, respectively. The identification method of the inner and outer boundaries of the penumbra is similar to that of \cite{Li+etal+2021}, and the threshold is fine-tuned to adapt better to the inner and outer boundaries of the penumbra on the continuum intensity images. Once more, the region grow method is used to determine the penumbra region. Finally, the hole filling method is used to remove the wrong identification region in the penumbra. Fig. \ref{dacay evolution} shows the identification boundaries of the inner (the red curves) and outer (the green curves) boundaries of the penumbra in continuum images based on the above process.

The unsigned magnetic flux ($\Phi$) is calculated by the following formula:$$ \Phi =\int \left |B_{R}\right |dA.$$where $dA$ represents the differential area.

The transverse magnetic field ($B_{t}$) is obtained by the following formula: $$B_{t}=\sqrt{B_{P}^{2}+B_{T}^{2} }.$$ 

The formula for calculating the magnetic field inclination angle ($\gamma$) is listed below: $$\gamma=\arctan( \frac{B_{t}}{\left | B_{R} \right |} )\frac{180^{\circ} }{\pi}.$$  

\section{Results}
\label{sect:result}
\begin{figure}[ht]    
	\setlength{\abovecaptionskip}{-0.8cm}
	\includegraphics[scale=0.8]{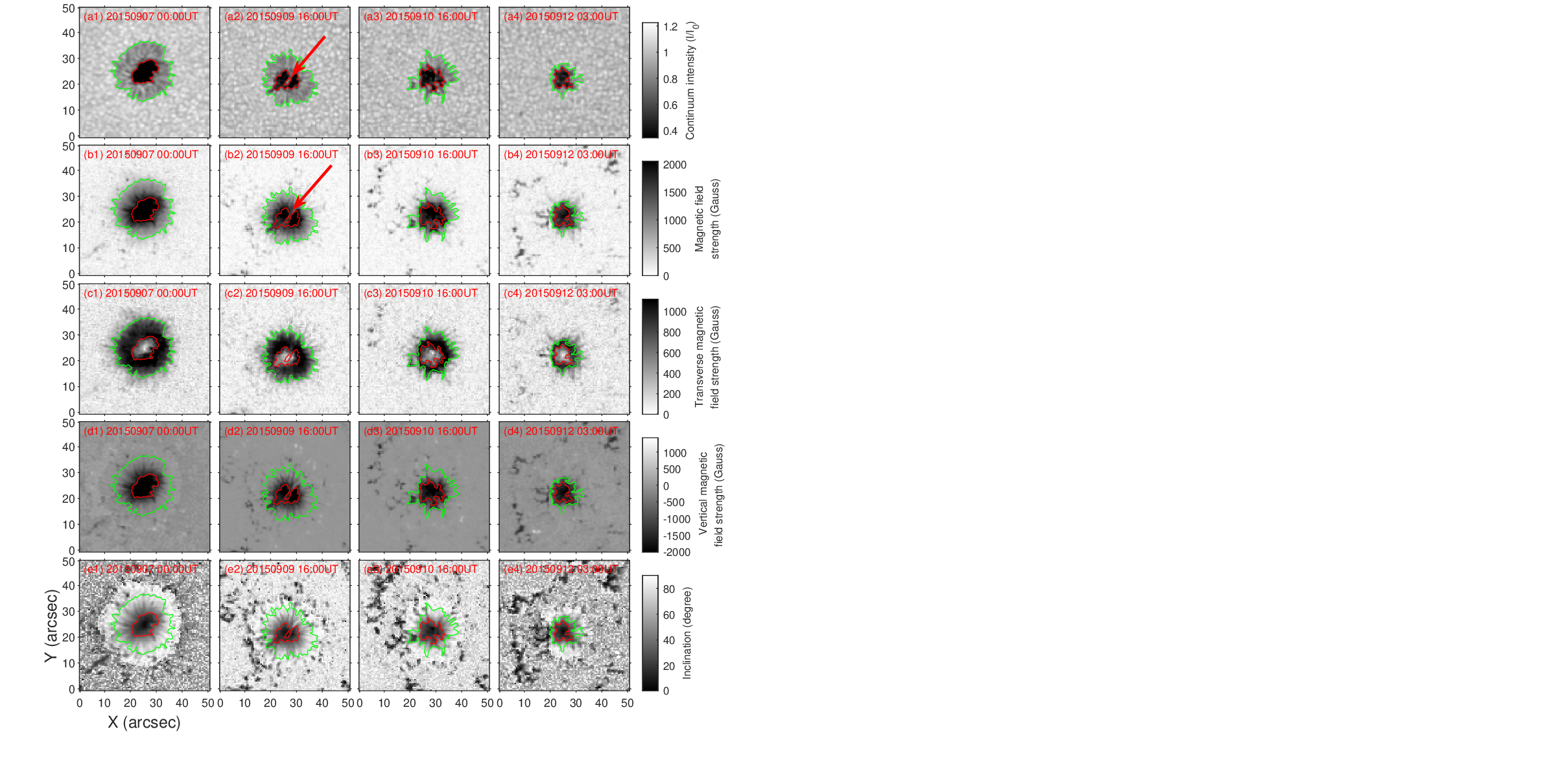}
	\caption{Evolution process of the sunspot in AR NOAA 12411. The inner and outer boundaries of the penumbra are depicted by the red contour at 0.55$I_{0}$ and green contour at 0.94$I_{0}$, respectively. The red arrow marks the bright bridge that appeared during the evolution of the sunspot. The upper most row shows the continuum image at four moments of the sunspot evolution. The second, third, fourth, and fifth rows are the magnetic field strength, horizontal magnetic field strength, vertical magnetic field strength and inclination angle maps at the corresponding moment of the continuum ones in the first row.}	
	\label{dacay evolution}
	\centering
\end{figure}

The decay process of the $\alpha$-type sunspot is shown in Fig. \ref{dacay evolution}. The continuum intensity, total magnetic field strength, horizontal and vertical magnetic field strength, the inclination angle maps are displayed in Fig. \ref{dacay evolution} from the top row to the bottom row. The polarity of the sunspot is negative. In the early stage of the sunspot decay, it owns an annular penumbra, a bright bridge gradually forms in the umbra (see panel a2), and the vertical magnetic field strength map (panel d2) show the vertical magnetic field of the bright bridge is weaker than that of the umbra. A lot of Moving Magnetic Features (MMFs) are seen around the sunspot in magnetic field strength maps. From the vertical magnetic field strength maps, a large number of MMFs come out of the sunspot. A part of a network gradually presents in panel d3. In the end, there is an obvious network around the sunspot (see panel d4). Some studies suggested that MMFs transported the magnetic flux of decaying sunspots to the surrounding network \citep{Deng+etal+2007, Verma+etal+2012}, and they are carried out of the decaying sunspot by moat flow \citep{ Harvey+Harvey+1973}. With the evolution of the sunspot, the average transverse magnetic field strength of the umbra is weaker than that of the penumbra. For the average vertical magnetic field strength, the umbra is stronger than the penumbra. After 6:00 UT on September 12, 2015, the sunspot shatters into several pieces. Since the program do not work well for identifying subsequent process after that time, the follow-up tracking is not continued.

\begin{figure}[]
	\centering
	\setlength{\abovecaptionskip}{-2.6cm}
	\includegraphics[scale=0.38]{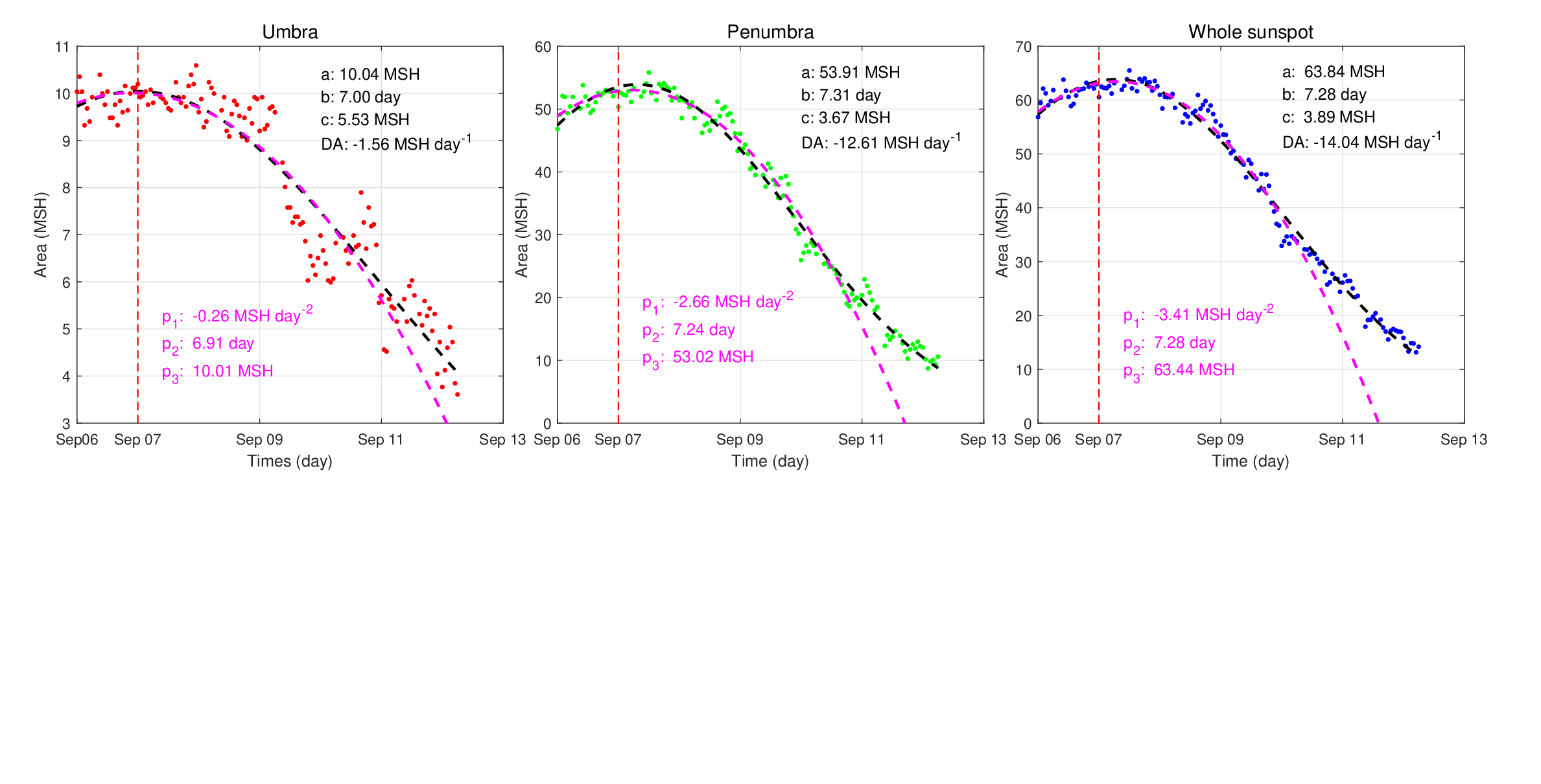}
	\caption{The evolution of the sunspot area. The left panel: the evolution of the area of the umbra with time. The middle panel: the evolution of the area of the penumbra with time. The right panel: The evolution of the area of the whole sunspot with time. The black curves represent Gaussian fitting to the area scatter plots of the different components (whole sunspot, umbra and penumbra) of the sunspot in the form: $A(t)=ae^{-(\frac{t-b}{c} )^{2} }$, $A(t)$ and $t$ represent the area of sunspot and time. $a$, $b$, $c$, $DA$ represent maximum area value, time starting point and standard deviation of the area decay process, and area decay rate (Unit: $\rm MSH\ day^{-1}$) of the corresponding sunspot region, respectively. The magenta curves represent the quadratic fitting (formula: $A(t)=p_{1}(t-p_{2})^{2}+p_{3}$) to the area scatter plots between the beginning of observation and the inflection point (a point whose the second derivative is equal to zero) on the right side of the symmetry axis of Gaussian fitting. The sign of the $p_{1}$, and the $p_{2}$, $p_{3}$ represent the opening direction of the fitting curve, the moment when the vertex of the quadratic curve is located, and the vertex value of the parabola, respectively.}
	\label{area}
\end{figure}

The decay of the sunspot area with time is shown in Fig. \ref{area}. From left to right, the area decay curves of the umbra, penumbra, and whole sunspot are shown in order. The abscissa of each panel represents time (Unit: day), and the ordinate represents the area of the corresponding region. Before 00:00 on September 7 (marked by the red vertical lines), the sunspot is positioned near the edge of the Sun, where the magnetic field is not accurately measured and the area is not affected. To show the complete sunspot decay, the area data from September 6 to September 7 are added. The area unit is set to millionths solar hemisphere (MSH), and 1 MSH=3.32 M$\rm m^{2}$. The observational data in the three panels take appearance of Gaussian profiles. A standard Gaussian function is used to fit the area as follows: $A(t)=ae^{-(\frac{t-b}{c} )^{2} }$, where $t$, $A(t)$, $a$, $b$, and $c$ represents time (day), area (MSH), maximum area value, starting time to decay and standard deviation of the decaying area, respectively. The function steepness at the median of maximum and minimum of the area is considered to be the decay rate of the area, marked as DA \citep{Murakozy+etal+2014}. All fitting parameter values are placed in each subgraph. To compare with previous studies which have shown quadratic decay of the area of the sunspot, a classical quadratic fitting is performed to the area data between the beginning of observation and the inflection point (a point whose the second derivative is equal to zero) on the right side of the symmetry axis of the Gaussian fitting, and the quadratic fitting curves are the magenta curves in the Fig. \ref{area}. The peak values for the decay of the area in the umbra, penumbra, whole sunpot determined by the quadratic fitting are 10.01 MSH, 53.02 MSH and 63.44 MSH, respectively. The quadratic fitting curves are in good agreement with the scatter plots in the early stages of the sunspot decay on different components of the sunspot, and there will be big deviations between them in the late stages of the sunspot decay. We take the moments at which the Gaussian fitting vertexes are located as the starting times for the decay of different sunspot components. The area of the umbra is about 9.70 MSH at the beginning of the observation, and increases to the peak (10.04 MSH), then decreases rapidly to 7 MSH. After that, it decreases slowly. The umbra has an area of 4.30 MSH at the end of observation. The area of the penumbra is 47 MSH at the beginning of obsevation, and increases to the peak (54 MSH), then quickly decreases to 30 MSH. Finally, it decays slowly to 8.70 MSH. For the whole sunspot, the area increases from 57 MSH to 63.84 MSH at the beginning, it starts to decay at 63.84 MSH, reduces rapidly to 35 MSH, and then slowly decays to the end of the observation. The area of the whole sunspot has a value of 12.30 MSH at the end of the observation. The Gaussian fitting results show the umbra begins to decay earlier than the penumbra from parameter $b$. \cite{Li+etal+2021} also found that, in some $\rm \alpha$-configuration sunspots, the umbra first starts to decay. The standard deviations (from parameter $c$) of the umbra and penumbra are 5.53 MSH and 3.67 MSH, implying that the penumbra decay process is more continuous. The area decay rates of the umbra, penumbra and whole sunspot are -1.56 $\rm MSH\ day^{-1}$, -12.61 $\rm MSH\ day^{-1}$ and -14.04 $\rm MSH\ day^{-1}$. The penumbra decays faster than the umbra.

\begin{figure}[htbp]
	\centering
	\setlength{\abovecaptionskip}{0.1cm}
	\includegraphics[scale=0.56]{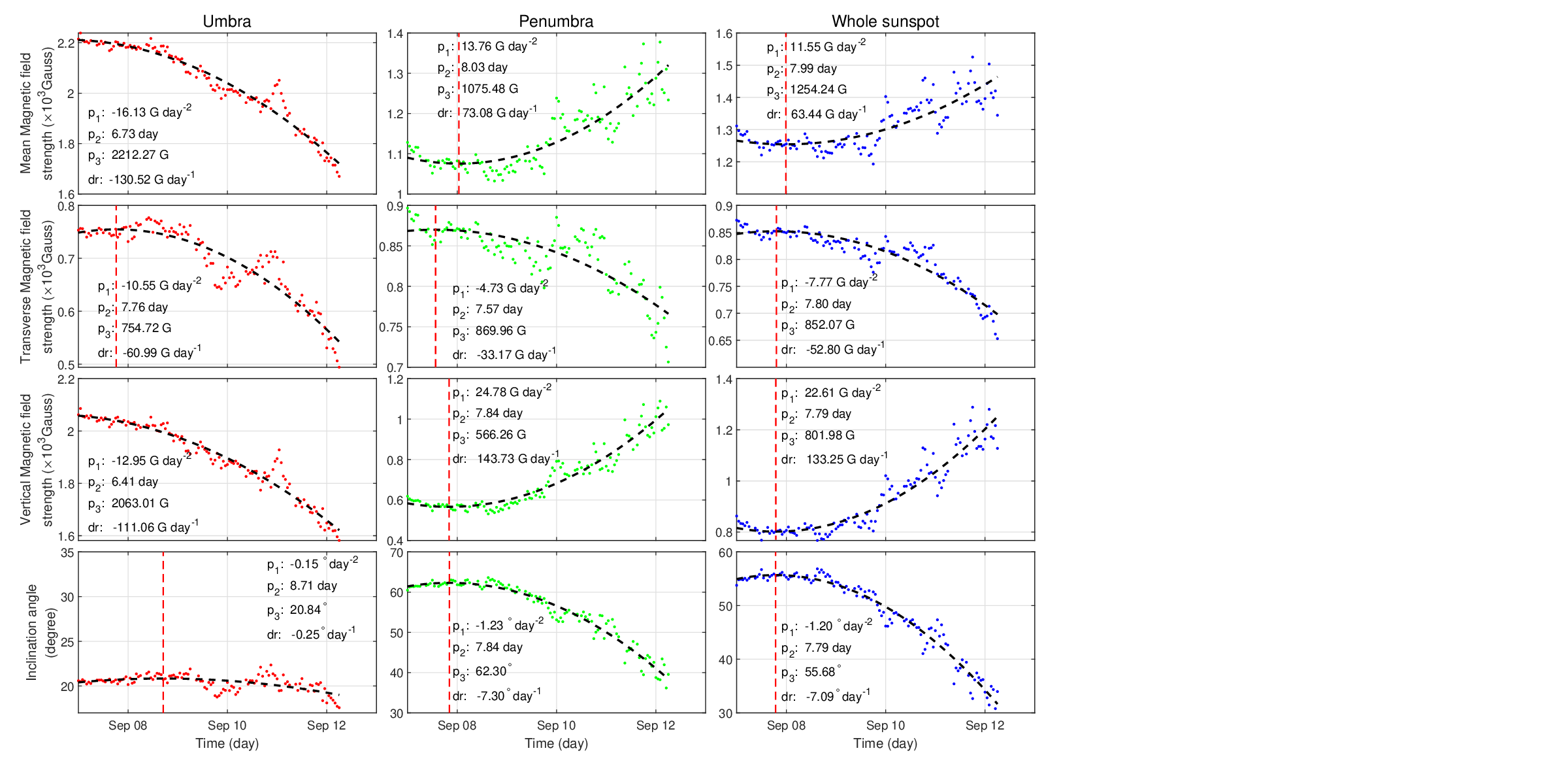}
	\caption{The evolution of sunspot magnetic field parameters. The first to fourth rows are the average magnetic field strength, transverse magnetic field strength, vertical magnetic field strength, and inclination angle, respectively. The red, green scatter plots represent respectively the corresponding magnetic field properties of the umbra and penumbra, respectively. The quadratic curve is used to fit the changing trends of the magnetic field parameters in the form: $M(t)=p_{1}(t-p_{2})^{2}+p_{3}$, $M(t)$ and $t$ are corresponding magnetic field parameter values (Unit: Gauss or degree) and time (Unit: day). The sign of the $p_{1}$, and the $p_{2}$, $p_{3}$ represent the opening direction of the fitting curve, the moment when the vertex of the quadratic curves is located, and the vertex value of the parabola, respectively. The red reference lines represent the moments at which the peaks of the quadratic fittings occur.}
	\label{magnetic field}
\end{figure}

Fig. \ref{magnetic field} shows the development of the magnetic field properties of the sunspot. The Quadratic curve is used to fit the changing trends of the magnetic field parameters in the form: $M(t)=p_{1}(t-p_{2})^{2}+p_{3}$. $M(t)$ and $t$ are corresponding magnetic field parameter values (Unit: Gauss or degree) and time (Unit: day), respectively. The sign of the $p_{1}$, and the $p_{2}$, $p_{3}$ represent the opening direction of the fitting curve, the moment when the vertex of the quadratic curve is located, and the vertex value of the parabola, respectively. The average change rate ($dr$) for each panel is defined as the steepness of the fitting curve at the median of the maxima and minima. Its unit is $\rm G\ day^{-1}$ or $\rm degree\ day^{-1}$. We describe the change of sunspot magnetic field parameters from the beginning of observation, and take the moments at which the peaks of the quadratic fitting curves as the starting times of the corresponding magnetic field parameter to decay. The red vertical lines represent the moments at which the peaks of the quadratic fittings curves. The mean magnetic field strength of the umbra is 2200 G at the beginning of the observation, then it decreases slowly first and rapidly decreases to 1720 G at the end. The mean magnetic field strength of the penumbra drops slowly from 1090 G to a minimum value of 1075 G first, it starts to increase slowly from 1075 G and then fast increases to 1380 G. The mean magnetic field change rate of the umbra and penumbra are -130.52 $\rm G\ day^{-1}$ and 73.08 $\rm G\ day^{-1}$, respectively. Evidently, the mean magnetic field strength change rate of the penumbra is smaller than that of the umbra. The transverse magnetic field of the umbra slowly increases from 750 G to 755 G in the initial stage, while that of the penumbra increases slowly from 867 G to 870 G. The transverse magnetic field strength for umbra and penumbra are 755 G and 870 G at the peaks, decrease slowly at the beginning and then rapidly. The transverse magnetic field strength for the umbra and the penumbra have values of 540 G and 765 G at the end. The maximum transverse magnetic field strength values for the umbra and the penumbra are 755 G and 870 G, respectively. The decay rates of the transverse magnetic field strength for the umbra and penumbra are -60.99 $\rm G\ day^{-1}$ and -33.17 $\rm G\ day^{-1}$, respectively. The transverse magnetic field strength of the umbra is weaker than that of the penumbra all the time,  and the transverse magnetic field strength of the umbra decay faster than that of the penumbra. The vertical magnetic field strength of the umbra starts with a value of 2063 G, decreases slowly first and then rapidly to 1620 G. On the other hand, the vertical magnetic field strength of the penumbra decreases slowly from 580 G to the valley (566 G) during the beginning stage of observation, and begins to increase slowly first and then rapidly to 1030 G. The change rate of the vertical magnetic field strength of the umbra and penumbra are respectively -111.06 $\rm G\ day^{-1}$ and 143.73 $\rm G\ day^{-1}$. Therefore, the vertical magnetic field strength of the umbra is rapidly decaying, while that of the penumbra is rapidly increasing. And furthermore, the vertical magnetic field value of the penumbra changes faster than that of the umbra. With the decrease of the mean magnetic field strength of the sunspot umbra, both the mean transverse magnetic field and mean vertical magnetic field of the umbra decrease. It suggests that the magnetic field of the umbra is gradually decreasing. The average inclination angle of the umbra varies slightly. The mean value and standard deviation of the inclination angle of the umbra are calculated as $20.36^{\circ} $ and $0.9^{\circ}$, respectively. It indicates the magnetic field inclination angle does not change significantly although the average magnetic field strength in the umbra decreases, and the magnetic field has been in a relatively vertical state. The penumbra inclination angle increases slowly from $60^{\circ}$ to its peak ($62^{\circ}$) , then decreases to $38^{\circ}$ with a decay rate of $\rm -7.30^{\circ}\ day^{-1}$. It clearly specifies that the penumbra magnetic field increases rapidly and becomes more vertical during the sunspot decay. During the process of decay, the variation trends of the magnetic field parameters in the whole sunspot are similar to that of the penumbra, since the area of the penumbra is much larger than that of the umbra.

\begin{figure}[htbp]
	\centering
	\setlength{\abovecaptionskip}{-2.8cm}
	\includegraphics[scale=0.39]{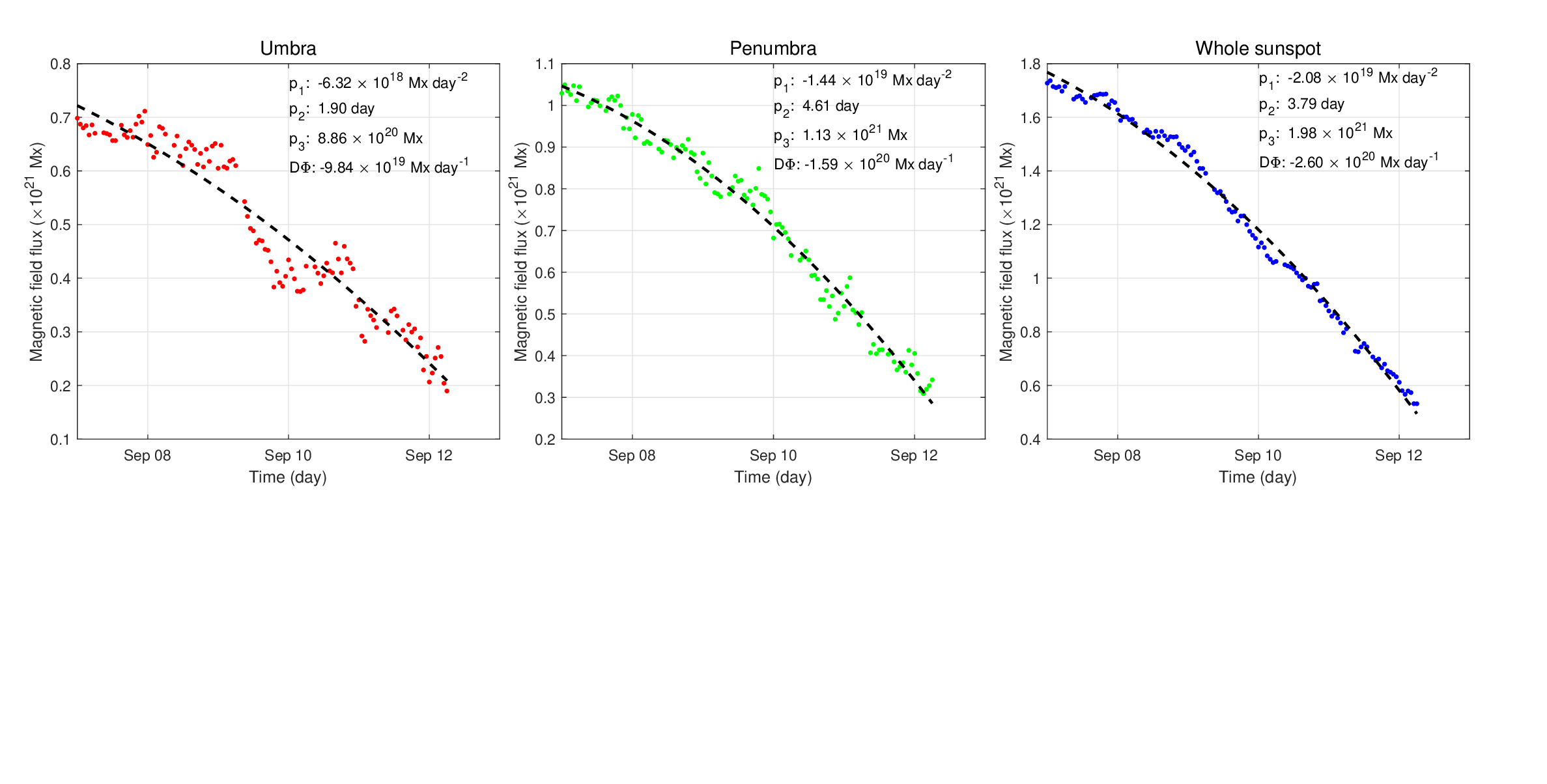}
	\caption{The evolution of sunspot magnetic flux over time. The left, middle, and right panel represent the evolution of the magnetic flux in the umbra, penumbra, and whole sunspot with time, respectively. The quadratic curves are used to fit the change trends of the magnetic flux in the three regions of the sunspot, expressed as: $\Phi(t)=p_{1}(t-p_{2})^2+p_{3}$, $\Phi(t)$ and $t$ are magnetic flux (Unit: Mx) and time (Unit: day). The sign of the $p_{1}$, and the $p_{2}$, $p_{3}$ evaluates the opening direction, the moment of maximum magnetic flux, and the vertex value of the quadratic curve, respectively. $D\Phi$ represents the decay rate of the magnetic flux and its unit is $\rm Mx\ day^{-1}$.}
	\label{flux}
\end{figure} 

The sunspot magnetic flux with time is shown in Fig. \ref{flux}. The magnetic flux in the three regions of the sunspot all show quadratic decay trends with time. The quadratic function is used to fit the change trend of the magnetic flux in the three regions of the sunspot, expressed as: $\Phi(t)=p_{1}(t-p_{2})^2+p_{3}$. $\Phi(t)$ is the magnetic flux (Unit: Mx) at time $t$ in unit of day, the sign of the $p_{1}$, and the $p_{2}$, $p_{3}$ evaluates the opening direction, the moment of maximum magnetic flux, and the vertex value of the quadratic curve, respectively. The steepness of the curve at the median of the maximum and minimum values is defined as the decay rate of the magnetic flux (marked as $D\Phi$ and its unit is $\rm Mx\ day^{-1}$). The magnetic flux of the umbra decreases slowly from $7.20\times10^{20}$ Mx at 00:00 UT on September 7, and then decreases fast to $2.10\times10^{20}$ Mx at 3:00 UT on September 12. At 00:00 UT on September 7, the magnetic flux of the penumbra is measured to be a value of $10.04\times10^{20}$ Mx. It decreases slowly at first, and then decrease fast to $2.90\times10^{20}$ Mx. The change of magnetic flux of the umbra is relatively discrete, while that of the penumbra is relatively continuous. The magnetic flux of the whole sunspot decreases from $17.60\times10^{20}$ Mx at the beginning of observation to $5.30\times10^{20}$ Mx at the end. Its process is still a slow decrease first and then a fast decrease. The decay rates of magnetic flux in the umbra, penumbra and whole sunspot are $-9.84\times10^{19}$ $\rm Mx\ day^{-1}$, $-1.59\times10^{20}$ $\rm Mx\ day^{-1}$, and $-2.60\times10^{20}$ $\rm Mx\ day^{-1}$, respectively. The magnetic flux of the penumbra decay faster than that of the umbra.

\section{Conlusions and Discussion}
\label{sect:conclusion}
The decay process of an $\alpha$-configuration sunspot has been studied. The quantitative evolution of the area, magnetic field properties, and the magnetic flux of the sunspot are calculated, respectively. The following main results are obtained:\\
   (1) The area decay of the sunspot in its umbra, penumbra, and whole sunspot can be described by the Gaussian-variation. The area decay rate of the umbra, penumbra and whole sunspot are -1.56 $\rm MSH\ day^{-1}$, -12.61 $\rm MSH\ day^{-1}$ and -14.04 $\rm MSH\ day^{-1}$, respectively.\\
   (2) With the decay of the sunspot, the total magnetic field strength and vertical magnetic field strength of the penumbra enhances, and the magnetic field of the penumbra becomes more vertical. Meanwhile, the magnetic field strength and vertical magnetic field strength for the umbra decays continuously. The magnetic field inclination angle of the umbra changes slightly and is around 20 degrees.\\
   (3) The magnetic flux decay of the sunspot in the umbra, penumbra, and whole sunspot behaves quadratically. The magnetic flux decay rate of the umbra, penumbra and entire region are $-9.84\times10^{19}$ $\rm Mx\ day^{-1}$, $-1.59\times10^{20}$ $\rm Mx\ day^{-1}$ and $-2.60\times10^{20}$ $\rm Mx\ day^{-1}$, respectively.
   
The previous studies have shown that the area decay process of the sunspot is linear \citep{Solanki+2003, Li+etal+2021} or quadratic \citep{Litvinenko+Wheatland+2015}. \cite{Murakozy+etal+2014} studied the area growth rates of 399 sunspot groups based on the solar and Heliospheric Observatory/Michelson Doppler Imager - Derecen Data (SDD) sunspot catalogue, and found that asymmetric Gaussian function well fitted the growth and decay process of these sunspot groups. \cite{Murak+2020} also studied the relationship between the decay rate of umbral and penumbral area and their total area based on SDD sunspot catalog, and concluded that asymmetric Gaussian function also fitted well the decay process of these sunspot groups. In our study, the area decay of sunspots is fitted well by Gaussian function, similar to that of \cite{Murakozy+etal+2014}. On the other hand, the quadratic fitting curves are in good agreement with the scatter plots in the early stages of sunspot decay, however, there are big deviations between them in the late stages of the sunspot decay. It indicates that the quadratic fitting may be better for the early decay stage of the sunspot, while the Gaussian fitting is better for the entire decay of the sunspot. The area decay rates of the umbra, penumbra, and whole sunspot are -1.56 $\rm MSH\ day^{-1}$, -12.61 $\rm MSH\ day^{-1}$ and -14.04 $\rm MSH\ day^{-1}$, respectively. This result is similar to the study of \cite{Li+etal+2021}, but slightly smaller than the research of \cite{Martinez+etal+1993}.

 \cite{Verma+etal+2018} studied penumbral decay of AR NOAA 12597 observed by GREGOR, and found that a darkened region whose magnetic field properties are similar to that of an umbra was observed in the penumbral filament. It indicated that the magnetic field in the penumbra became more vertical due to flux emergence. \cite{Wang+etal+2004} researched a rapid penumbral decay following three flares and observed that the magnetic field in penumbra became more vertical. They suggested that magnetic field of part of the penumbra was converted into magnetic field state of the umbra. By studying the penumbra of several delta sunspots, \cite{Wang+etal+2012} found that the magnetic field of the penumbra tended to be more vertical during flares. It was attributed to the magnetic field reconstruction related to the flares. In our observation, the average magnetic field of the penumbra becomes stronger, and the magnetic field becomes more vertical. However, the magnetic field strength of the umbra shows a decline, and its inclination angle keeps a constant. Maybe due to the submergence of the horizontal magnetic field in the penumbra during sunspot decay \citep{Rempel+2015}, while relatively vertical magnetic field still leaves in the photosphere, and thus the magnetic field in the penumbra becomes more vertical on average. The horizontal magnetic field carries more brightness and heat than the vertical magnetic field, and when the horizontal magnetic field is lost in the penumbra, the penumbra will either become dark or the convection within a more vertical magnetic field which resemble regular granulation will dominate the penumbra (e.g. around pores). The observations indicates that the penumbra may be transformed into umbra during sunspot decay.
 
 The decay process of the umbral magnetic flux is relatively discrete, and exhibits a quadratic trend. The decay process of the penumbral magnetic flux also takes appearance of quadratic profile, which is consistent with \cite{Benko+etal+2021}. The decay rates of magnetic flux in the umbra, penumbra and whole sunspot are $-9.84\times10^{19}$ $\rm Mx\ day^{-1}$, $-1.59\times10^{20}$ $\rm Mx\ day^{-1}$, and $-2.60\times10^{20}$ $\rm Mx\ day^{-1}$. The decay rate of magnetic flux in whole sunspot is consistent with the result of \cite{Sheeley+etal+2017}. In the process of the sunspot decay, there will be a lot of moving magnetic features (MMFs) coming out from sunspot. The magnetic flux decay of the sunspot may be attributed to MMFs. The previous researches suggest that the magnetic flux in the sunspot is transported by MMFs to the surrounding network during sunspot decay \citep{Verma+etal+2012, Deng+etal+2007}. Some scholars believe that the MMF is a prolongation of the evershed flow in the penumbra \citep{Solana+etal+2006, Schlichenmaier+2002}.

\begin{acknowledgements}
We wish to express our gratitude to
	the anonymous referee for constructive comments and suggestions. We thank the SDO/HMI teams for the active region data with stray light correction support. This work is supported by the National Natural Science Foundation of China (11973084,11873087,12003066, 11527804), Yunnan Key Laboratory of Solar Physics and Space Science under number 202205AG070009, Yunnan Science Foundation of China under numbers: 202201AT070194, Youth Innovation Promotion Association, CAS (Nos 2019061), CAS “Light of West China” Program, Yunnan Science Foundation for Distinguished Young Scholars No. 202001AV070004, Key Research and Development Project of Yunnan Province under number 202003AD150019,and Young Elite Scientists Sponsorship Program by YNAST.9.
	
\end{acknowledgements}
\bibliographystyle{raa}
\bibliography{msRAA-2022-0272}

\end{document}